\documentclass{ws-ijmpd}
\usepackage{amssymb}
\usepackage{amsmath}
\usepackage{graphicx}
\usepackage{epsfig}
\usepackage{bm}
\begin{document}

\markboth{Sergei M. Kopeikin}
{Gravitomagnetism and the Speed of Gravity}

%
\catchline{}{}{}{}{}
%

\title{Gravitomagnetism and the Speed of Gravity}

\author{Sergei M. Kopeikin}

\address{Department of Physics and Astronomy, University of Missouri-Columbia, \\Columbia, Missouri 65211, USA\\
kopeikins@missouri.edu}

\maketitle


\begin{abstract}\noindent
Experimental discovery of the gravitomagnetic fields generated by translational and/or rotational currents of matter is one of primary goals of modern gravitational physics. The rotational ({\it intrinsic}) gravitomagnetic field of the Earth is currently measured by the Gravity Probe B. The present paper makes use of a parametrized post-Newtonian (PN) expansion of the Einstein equations to demonstrate how the {\it extrinsic} gravitomagnetic field generated by the translational current of matter can be measured by observing the relativistic time delay caused by a moving gravitational lens. We prove that measuring the {\it extrinsic} gravitomagnetic field is equivalent to testing relativistic effect of the aberration of gravity caused by the Lorentz transformation of the gravitational field. We unfold that the recent Jovian deflection experiment is a null-type experiment testing the Lorentz invariance of the gravitational field (aberration of gravity), thus, confirming existence of the {\it extrinsic} gravitomagnetic field associated with orbital motion of Jupiter with accuracy 20\%. We comment on erroneous interpretations of the Jovian deflection experiment given by a number of researchers who are not familiar with modern VLBI technique and subtleties of JPL ephemeris. We propose to measure the aberration of gravity effect more accurately by observing gravitational deflection of light by the Sun and processing VLBI observations in the geocentric frame with respect to which the Sun is moving with velocity $\sim$ 30 km/s. 
\end{abstract}

\section{Introduction}

A gravitomagnetic field, according to Einstein's theory of general relativity, arises from moving matter (mass current) just as an ordinary magnetic field arises in Maxwell's theory from moving charge (electric current). The weak-field linearized theory of general relativity unveils a mathematical structure comparable to the Maxwell equations \cite{bct,m1,rugt}.  This weak-field approximation splits gravitation into components similar to the electric and magnetic field. In the case of the gravitational field, the source is the mass of the body, whereas in the case of the electromagnetic field, the source is the charge of the particle.  Moving the charge particle creates a magnetic field according to Amp\`ere's law. Analogously, moving the mass creates a mass current which generates a gravitomagnetic field according to Einstein's general relativity.  Amp\`ere-like induction of a gravitomagnetic field (gravitomagnetic induction) in general relativity has been a matter of peer theoretical study since the Lense-Thirring paper \footnote{English translation of the original Lense-Thirring paper is given by Mashhoon et al. in \cite{mht}. More recent derivations of the Lense-Thirring effect on the orbital motion of a test particle can be found in \cite{ashall,ior}. For detailed study of the Lense-Thirring effect caused by the rotational currents of matter see the textbook \cite{cw}.}  Nowadays this problem can be tackled experimentally.

There are two types of mass currents in gravity. The first type is brought about by the rotation of matter around body's center of mass. It generates {\it intrinsic} gravitomagnetic field tightly associated with body's angular momentum (spin) and so far the basic research in gravitomagnetism has been focused on the discussion of its various properties \cite{cw}. The second type of the mass current is caused by translational motion of matter. It generates {\it extrinsic} gravitomagnetic field that depends on the frame of reference of observer and can be completely eliminated in the rest frame of matter. This property of the {\it extrinsic} gravitomagnetic field is a direct consequence of the Lorentz invariance of Einstein's gravity field equations \cite{ashss} and its experimental testing is as important as that of the {\it intrinsic} gravitomagnetic field. The point is that both the {\it intrinsic} and the {\it extrinsic} gravitomagnetic fields represent actually different sides of a single gravitomagnetic field associated with $g_{0i}$ component of the metric tensor which can be split in pure curl (intrinsic) and gradient (extrinsic) parts. Furthermore, detection of the {\it extrinsic} gravitomagnetic field probes the Lorentz invariance of the gravitational field which determines its causal nature, that is whether the fundamental speed limit $c_g$ for gravitational field (= the speed of gravity) is the same as the fundamental speed limit $c$ for electromagentic field (= the speed of light). Experimental measurement of the {\it intrinsic} gravitomagnetic field of the rotating Earth is currently under way by the Gravity Probe B mission \cite{gpb} with the expected accuracy of 1\% or better \cite{gpb1}. Laser ranging of LAGEOS and other geodetic satellites verify its existence as predicted by Einstein's general relativity \cite{iormor,cp} although the claimed accuracy is rather controversial \cite{ior2}\footnote{Other interesting proposals for measuring the {\it intrinsic} gravitomagnetism in the Solar system are described in \cite{ior2,ior3,ior4}.} 

The goal of the present paper is to show that the {\it extrinsic} gravitomagnetic field can be measured in high-precision relativistic time-delay experiments where light (radio wave) interacts with the gravitational field of a moving massive body (gravitational lens) \footnote{Ashby and Sahid-Saless \cite{ashss} discussed a similar idea for measuring the {\it extrinsic} gravitomagnetic field by observing geodetic precession of a satellite's orbit caused by orbital motion of a massive body with respect to the local inertial frame of the satellite.}. Relativistic light deflection in the rest (static) frame of the light-ray deflecting body was calculated in optics by Einstein \cite{ein} and Shapiro \cite{shapiro} had derived the relativistic time delay in radio. General relativistic problem of the gravitational deflection and delay of light by an arbitrary moving body has been solved in our papers \cite{ks,km} which did not discern relativistic effects associated with the Lorentz-invariant properties of light and gravitational field, that is the metric tensor $g_{\alpha\beta}$, the affine connection $\Gamma^\alpha_{\beta\gamma}$, and the curvature tensor \cite{LL,mtw}. In the present paper we use a standard PN expansion of general relativistic equations parametrized by a {\it speed-of-gravity} parameter $\epsilon$ that allows us to clearly separate the relativistic effects associated with the Lorentz boosts of the Einstein gravity field equations from that of the Maxwell equations. Formally, the speed-of-gravity parametrization in Einstein's equations is achieved after replacing matter's current ${\bm j}$, velocity of the matter ${\bm v}$, and {\it all} partial time derivatives of the metric tensor in accordance to the following rules (see \cite{fs,cqg} for more detail)
\begin{equation}
\label{jtd}
{\bm j}\rightarrow\epsilon{\bm j}\;,\qquad\qquad{\bm v}\rightarrow\epsilon{\bm v}\;,\qquad\qquad\frac{\partial }{\partial t}\rightarrow\epsilon\frac{\partial }{\partial t}\;. 
\end{equation}
It is apparent that the replacements (\ref{jtd}) do not change anything in the structure of the general relativistic equations because it is fully equivalent \cite{fs,cqg} to introduction to these equations of the {\it dynamic} time coordinate $x^0=c_g t$, where $c_g\equiv c/\epsilon$. One should not think that the introduction of the parameter $c_g$ leads to the replacement of the Minkowski metric $\eta_{\alpha\beta}={\rm diag}(-1,+1,+1,+1)$ to another one $f_{\alpha\beta}={\rm diag}(-c^2/c_g^2,+1,+1,+1)$. The Minkowski metric $\eta_{\alpha\beta}={\rm diag}(-1,+1,+1,+1)$ of the background flat space-time remains fixed because we work with one and the same value of the time coordinate $x^0$. What changes is the character of the transformation of the field variables (the metric tensor perturbation $h_{\alpha\beta}$ and the affine connection $\Gamma^\alpha_{\;\mu\nu}$) residing on the background when one goes from one reference frame to another.

Parameter $\epsilon$ singles out the Lorentz group of transformation of the gravitational field given by the matrix $\Lambda^\alpha_\beta(\epsilon)$ depending on the parameter $\epsilon$. Explicit dependence of this matrix on parameter $\epsilon$ makes it look formally different from the Lorentz group of transformation of electromagnetic field, described by the matrix $\lambda^\alpha_\beta\equiv\Lambda^\alpha_\beta(\epsilon=1)$, so that the effects associated with null characteristics of light and those of gravity can not be confused. As long as one considers orbital motion of massive bodies (artificial satellites, planets, binary pulsar, etc.) the difference between $c_g$ and $c$ shows up only in the relativistic effects of order $(v/c)^2$ and higher, where $v$ is a characteristic velocity of the massive body. However, observing propagation of light in the gravitational field of a massive moving body makes the difference between $c_g$ and $c$ to be measurable already in the linear terms of order $v/c$ due to the fundamental nature of the interaction of the electromagnetic field with the affine connection in the equations of light geodesics. 

The {\it speed-of-gravity} parameter $\epsilon$ is more general than the parameters $\alpha_1, \alpha_2, \alpha_3$ of the PPN formalism \cite{will}. This is because these parameters describe violation of the Lorentz symmetry of the metric tensor alone. Linear space-time transformation of the metric tensor tests the Lorentz invariance of the gravitational field locally at a single point of the space-time manifold. The affine connection and curvature of the manifold are formed from the metric tensor derivatives which must transform in accordance with the group of the Lorentz transformations of the metric, that is $\Lambda^\alpha_\beta(\epsilon)$. Therefore, to preserve this property parameter $\epsilon$ must be introduced in front of the time derivatives of the metric tensor (see equation (\ref{jtd})) that is in the affine connection and the curvature. By construction the Lorentz violation of the affine connection and the curvature in the PPN formalism are due to the broken symmetries of the metric tensor alone \cite{will}. PPN formalism postulates \cite{will} that the metric tensor derivatives are transformed in accordance with the group of the Lorentz transformation of electromagnetic field. However, according to Einstein \cite{LL,mtw} and results of all previous gravitational experiments \cite{will-r} the metric tensor describes gravitational field alone and, hence, its derivatives must be normalized to the speed of gravity $c_g$ and transformed in accordance with the Lorentz group of the gravitational sector of the theory. According to Einstein \cite{LL,mtw} the Lorentz groups of the transformation of both gravitational and electromagnetic fields must be identical but this theoretical foresight must be tested experimentally. The compatibility of the two groups can be tested if one performs a gravitational experiment in which both gravitational and electromagnetic fields are transformed from one inertial frame to another. This kind of experiments can be presently done with VLBI observations by observing propagation of light in the gravitational field of a moving massive body as demonstrated in \cite{k1,fk}. 

In the rest of this paper we derive the $\epsilon$-parametrized PN formula for the general-relativistic time delay and apply it for analysis of the relativistic effect of the {\it retardation of gravity} which has been measured recently in the Jovian experiment \cite{k1,fk}. We show that in the framework of the chosen $\epsilon$-parameterization the observed effect can be interpreted as: (1) measurment of the {\it extrinsic} gravitomagnetic field of Jupiter caused by its orbital motion, (2) verification that gravity is Lorentz invariant, (3) proof that the fundamental speed $c_g$ for gravity does not exceed the speed of light $c$ in vacuum. Finally, we discuss a new VLBI experiment in the field of the Sun in which the Lorentz invariance of gravity and the equality $c_g=c$ can be measured with accuracy up to 1\%.       

\section{Gravitomagnetism} 

Our parametrization scheme of the Einstein equations works in any approximation \cite{cqg}. For the sake of simplicity 
we shall work with the linearized PN equations of general relativity which is sufficient for adequate treatment of the light-ray deflection experiment. The PN expansion is done with respect to parameter $\epsilon=c/c_g$, where $c_g$ is the fundamental speed limit for gravity. If $c_g\rightarrow\infty$ (that is $\epsilon\rightarrow 0$), the general relativity collapses to the instantaneous speed-of-gravity limit where all gravitomagnetic phenomena vanish and the gravitational interaction is immediate.

In the first PN approximation the metric tensor perturbation $h_{\alpha\beta}=g_{\alpha\beta}-\eta_{\alpha\beta}$ of the gravitational field contains both the gravitoelectric, $\Phi=(c^2/2)h_{00}$ and the gravitomagentic, $A_i=-(c^2/4)h_{0i}$ potentials \cite{bct,m1,rugt}. The space-space components of the metric tensor perturbation are $h_{ij}=(2/c^2)\Phi\delta_{ij}$. The gravitoelectric, ${\bm E}$, and the gravitomagnetic, ${\bm B}$, fields are defined as 
\begin{eqnarray}  
\label{1}
{\bm E}&=&-{\bm\nabla}\Phi-\frac{\epsilon}{c}\frac{\partial{\bm A}}{\partial t}\;,\\
\label{2}
{\bm B}&=&{\bm\nabla}\times{\bm A}\;,
\end{eqnarray}
where ${\bm\nabla}\equiv\partial/\partial_i$ is a spatial gradient and parameter $\epsilon$ is formally introduced in accordance to Eq. (\ref{jtd}). After imposing the de-Donder gauge condition 
\begin{equation}
\label{3}
\frac{\epsilon}{c}\frac{\partial\Phi}{\partial t}+{\bm\nabla}\cdot{\bm A}=0\;,
\end{equation}
the Einstein equations lead to the system of the gravitomagnetic field equations
\begin{eqnarray}
\label{4}
{\bm\nabla}\cdot{\bm E}&=&4\pi G\rho\;,\\
\label{5}
{\bm\nabla}\cdot{\bm B}&=&0\;,\\
\label{6}
{\bm\nabla}\times{\bm E}&=&-\frac{\epsilon}{c}\frac{\partial{\bm B}}{\partial t}\;,\\
\label{7}
{\bm\nabla}\times{\bm B}&=&\frac{\epsilon}{c}\frac{\partial{\bm E}}{\partial t}+\frac{4\pi G\epsilon}c{\bm j}\;,
\end{eqnarray}
where $\rho$ and ${\bm j}$ are the mass-density and mass-current of the gravitating matter defined in terms of the energy-momentum tensor $T_{\alpha\beta}$ of matter as $\rho=(T^{00}+T^{ii})/(2c^2)$ and $j^i=T^{0i}/c$ respectively \cite{bct,m1,rugt,LL,mtw}. 

By making use of Eqs. (\ref{1}), (\ref{3}), (\ref{4}) and (\ref{7}) one can derive the wave equations for the potentials
\begin{eqnarray}
\label{a1}
\left(-\frac{\epsilon^2}{c^2}\frac{\partial^2}{\partial t^2}+\nabla^2 \right)\Phi&=&-4\pi G\rho\;,\\
\label{a2}
\left(-\frac{\epsilon^2}{c^2}\frac{\partial^2}{\partial t^2}+\nabla^2 \right){\bm A}&=&-\frac{4\pi G\epsilon}c{\bm j}\;.
\end{eqnarray}
These equations are of the hyperbolic type and describe propagation of gravitational field with the speed $c_g=c/\epsilon$. In the case of the instantaneous speed-of-gravity theory $\epsilon=0$, Eqs. (\ref{a1}),(\ref{a2}) are reduced to a single Laplace-type equation for the scalar potential $\Phi$ and both the gravitomagnetic potential, ${\bm A}$, and the gravitomagnetic field, ${\bm B}$, vanish.

\section{The Lorentz Invariance and Aberration of Gravity} 

Let us neglect the intrinsic rotation of a light-ray deflecting body (Jupiter, Sun). This is because the current light deflection experiments are not sensitive enough to measure the deflection of light caused by the body's spin \footnote{See however publications \cite{km,ckmr,tara1,ser,lat,astrod} where the measurement of the {\it intrinsic} gravitomagnetic field in light-ray deflection experiments is discussed. We notice that recent proposal by Tartaglia and Ruggiero \cite{tr1,tr2} for measuring gravitomagnetic field in binary pulsars is impractical since it is completely absorbed by the, so-called, bending effect predicted in \cite{dk} and further discussed in \cite{wk}}. Our assumption eliminates from Eqs. (\ref{4})--(\ref{a2}) all currents associated with motion of matter around body's center of mass. Therefore, the {\it intrinsic} gravitomagnetic field of the light-ray deflecting body caused by its rotation will not be discussed in the rest of the present paper.    

Gravity field equations (\ref{a1}), (\ref{a2}) are invariant with respect to the Lorentz group with the matrix of transformation $\Lambda^\alpha_\beta(\epsilon)$ having a standard form \cite{LL,mtw}
\begin{eqnarray}
\label{mat1}
\Lambda^0_{\;0}(\epsilon)&=&\gamma_\epsilon\equiv\left(1-\beta_\epsilon^2\right)^{-1/2}\;,\\
\label{mat2}
\Lambda^0_{\;i}(\epsilon)&=&\Lambda^i_{\;0}(\epsilon)=-\gamma_\epsilon\beta_\epsilon^i\;,\\
\label{mat3}
\Lambda^i_{\;j}(\epsilon)&=&\delta^{ij}+\left(\gamma_\epsilon-1\right)\frac{\beta_\epsilon^i\beta_\epsilon^j}{\beta_\epsilon^2}\;,
\end{eqnarray}
where the boost parameter $\beta_\epsilon^i=\epsilon v^i/c$, and ${\bm v}$ is the orbital velocity of the gravitating body (Jupiter, Sun). The matrix of the inverse transformation $\bar\Lambda^\alpha_\beta(\epsilon)=\Lambda^\alpha_\beta(-\epsilon)$. If $\epsilon\not=1$, the Lorentz transformation of the gravitational field is different from that of the electromagentic field. Physically it means that weak gravitational waves must propagate along the null cone which is different from the null cone for electromagnetic signals. In the case of the instantaneous speed-of-gravity theory, when $\epsilon=0$, the transformation (\ref{mat1})--(\ref{mat3}) degenerates and the space-time manifold splits in absolute time and absolute space with a single gravitational potential $\Phi$ residing on it.

Equations (\ref{4})--(\ref{a2}) solved in the static frame of the light-ray deflecting body (Jupiter, Sun) show that the potentials are: $\Phi'=(c^2/2)h'_{00}=-GM/c^2r'$, ${\bm A}'=-(c^2/4)h'_{0i}=0$, and the fields: ${\bm E}'=-{\bm\nabla}\Phi'\not=0$, ${\bm B}'={\bm\nabla}\times{\bm A}'=0$, where $r'$ is the radial coordinate of a field point in the static frame. Transformation of the static field potentials to the moving frame is obtained from the metric tensor Lorentz  transformation 
\begin{equation}
\label{arun}
h_{\alpha\beta}=\Lambda^\mu_{\;\alpha}(-\epsilon)\Lambda^\nu_{\;\beta}(-\epsilon)h'_{\mu\nu}\;,
\end{equation}
and is given more explicitly by equations
\begin{eqnarray}
\label{b1}
\Phi&=&\gamma^2_\epsilon\left[\left(1+\beta_\epsilon^2\right)\Phi'+4\left({\bm\beta}_\epsilon\cdot{\bm A}'\right)\right]\;,\\
\label{b2}
{\bm A}&=&\gamma_\epsilon{\bm A}'+\gamma^2_\epsilon\left[\Phi'+\frac{2\gamma_\epsilon+1}{\gamma_\epsilon+1}\left({\bm\beta}_\epsilon\cdot{\bm A}'\right)\right]{\bm\beta}_\epsilon\;.
\end{eqnarray}
Comparison with PPN metric \cite{will} shows that our approach yields the following values of the PPN parameters \cite{cqg}: $\gamma=\beta=1$, $\alpha_1=8(\epsilon-1)$, $\xi=\zeta_1=\zeta_2=\zeta_3=\zeta_4=\alpha_2=\alpha_3=0$. We emphasize that measurement of $\epsilon$ is not equivalent to an independent measurement of $\alpha_1$ in the PPN formalism framework \cite{will-r}. Measurement of $\alpha_1$ is not possible without making an (artificial) assumption about existence of a preferred frame usually associated with the isotropy of the cosmic microwave background radiation \cite{will-r}. Measurment of $\epsilon$ does not rely upon this assumption and is fulfilled through the measurement of the retardation of gravity effect as shown in \cite{cqg,k1,fk}.

The Lorentz transformation (\ref{b2}) generates the gravitomagnetic potential ${\bm A}$ and the gravitomagnetic field ${\bm B}$ associated with the orbital motion of the light-ray deflecting body with respect to the barycentric frame of the solar system. Neglecting the quadratic terms in Eqs. (\ref{b1}), (\ref{b2}) one obtains in the linear approximation that in the barycentric frame of the solar system
\begin{equation}
\label{8}
\Phi=-\frac{GM}{c^2r}+O\left(\epsilon^2\right)\;,\qquad\qquad{\bm A}=\frac{\epsilon}{c}\Phi{\bm v}+O\left(\epsilon^2\right)\;,
\end{equation}
where $r=|{\bm r}|$, ${\bm r}={\bm x}-{\bm z}(t)$, and ${\bm z}(t)$ is a coordinate of the body's center of mass in the barycentric frame. We emphasize that the Lorentz transformations (\ref{b1}), (\ref{b2}) assume that at each instant of time the motion of the body is considered as a straight line with constant velocity ${\bm v}=d{\bm z}/dt$ taken at time $t_{\rm A}$ which can be chosen arbitrary since we neglect acceleration of the body.

The Lorentz transformation of the static field ${\bm E}$ of the body generates the gravitomagentic field
\begin{equation}
\label{8a}
{\bm B}=\frac{\epsilon}c\left({\bm v}\times{\bm E}\right)\;,
\end{equation}
in accordance with its definition (\ref{2}) and Eq. (\ref{8}).
The Lorentz transformations (\ref{b1}), (\ref{b2}) describe changes (aberrations) in the structure of the gravitational field of the body, measured in two different frames. Equations (\ref{8}), (\ref{8a}) describe this {\it aberration of gravity} effect in 1.5 PN approximation which is linear with repsect to $v/c_g$ like the aberration of light is linear with respect to $v/c$.

In classical electrodynamics a uniformly moving charge generates magnetic field. This is because the Lorentz transformation generates electric current which produces the magnetic field. The resulting magnetic field is real and can be measured. Its observation confirms that electromagentic field is Lorentz-invariant and its speed of propagation is $c$ \cite{jackson}. 
Similarly, the gravitomagnetic potential (\ref{8}) and the gravitomagnetic field (\ref{8a}) are real and lead to observable effects which can be measured in gravitational experiments, thus providing a test of the Lorentz invariance of gravitational field and measurement of its fundamental speed limit $c_g=c/\epsilon$ that must be equal $c$ in general relativity.

\section{Gravitomagnetism and Time Delay} 

Equations of light propagation in vacuum are null geodesics. The unperturbed light particle moves with the velocity $c$ in any frame. Post-Newtonian equation of light propagation parameterized by coordinate time $t$ reads \cite{cqg}
\begin{equation}
\label{lg}
\frac{d^2x^i}{ dt^2}=c^2k^\mu k^\nu\left(k^i \Gamma^0_{\mu\nu}-\Gamma^i_{\mu\nu}\right)\;,
\end{equation}
where $k^\mu=(1,{\bm k})$ is a null four-vector, and ${\bm k}=k^i$ is the unit Euclidean vector being tangent to the unperturbed photon's trajectory. The affine connection  $\Gamma'^\alpha_{\mu\nu}$ in a static frame is given by equations
\begin{eqnarray}
\label{ss2}
\Gamma'^0_{00}&=&\Gamma'^0_{ij}=\Gamma'^i_{0j}=0\;,\\
\label{ss3}
\Gamma'^0_{0i}&=&\Gamma'^i_{00}=-\frac{\partial\Phi'}{\partial x'^i}\;,\\
\label{ss7}
\Gamma'^i_{jp}&=&-\delta_{jp}\frac{\partial\Phi'}{\partial x'^i}+\delta_{ip}\frac{\partial\Phi'}{\partial x'^j}+
\delta_{ij}\frac{\partial\Phi'}{\partial x'^p}\;,
\end{eqnarray}
where $\Phi'=(c^2/2)h'_{00}=-GM/c^2r'$, and $r'=|{\bm x}'|$. The affine connection is associated in general relativity with the gravitational force and its transformation from the static to a moving frame must be done with the matrix $\Lambda^\alpha_{\;\beta}(\epsilon)$, that is
\begin{equation}
\label{xru}
\Gamma^\alpha_{\beta\gamma}=\Lambda^\alpha_{\;\sigma}(\epsilon)\Lambda^\mu_{\;\beta}(-\epsilon)\Lambda^\nu_{\;\gamma}(-\epsilon)\Gamma'^\sigma_{\mu\nu}\;.
\end{equation}
Substituting matrix of the Lorentz transformation (\ref{mat1})--(\ref{mat3}) to equation (\ref{xru}) and making use of equations (\ref{ss2})--(\ref{ss7}) along with (\ref{b1}),(\ref{b2}) yields
\begin{eqnarray}
\label{bb2}
\Gamma^0_{00}&=&-\frac{\epsilon}c\frac{\partial\Phi}{\partial t}\;,\\
\label{bb3}
\Gamma^0_{0i}&=&-\frac{\partial\Phi}{\partial x^i}\;,\\
\label{bb4}
\Gamma^0_{ij}&=&+2\left(\frac{\partial A^j}{\partial x^i}+
\frac{\partial A^i}{\partial x^j}\right)+\frac{\epsilon}c\frac{\partial\Phi}{\partial t}\delta_{ij}\;,\\
\label{bb5}
\Gamma^i_{00}&=&-
\frac{\partial\Phi}{\partial x^i}-4\frac{\epsilon}c\frac{\partial A^i}{\partial t}\;,\\
\label{bb6}
\Gamma^i_{0j}&=&-2\left(
\frac{\partial A^i}{\partial x^j}-\frac{\partial A^j}{\partial x^i}\right)+\frac{\epsilon}c\frac{\partial\Phi}{\partial t}\delta_{ij}\;,\\
\label{bb7}
\Gamma^i_{jp}&=&-\delta_{jp}\frac{\partial\Phi}{\partial x^i}+\delta_{ip}\frac{\partial\Phi}{\partial x^j}+
\delta_{ij}\frac{\partial\Phi}{\partial x^p}\;,
\end{eqnarray}
where the parameter $\epsilon$ appears explicitly in front of the time derivatives present in equations (\ref{bb2}), (\ref{bb4})--(\ref{bb6}) in agreement with the parametrization-of-gravity rule (\ref{jtd}). 

It is worthwhile to compare equations (\ref{bb2})--(\ref{bb7}) with the definition of the affine connection adopted in the PPN formalism \cite{will} which postulates 
\begin{eqnarray}
\label{pp2}
\Gamma^0_{00}&=&-\frac{1}c\frac{\partial\Phi}{\partial t}\;,\\
\label{pp3}
\Gamma^0_{0i}&=&-\frac{\partial\Phi}{\partial x^i}\;,\\
\label{pp4}
\Gamma^0_{ij}&=&+2\left(\frac{\partial A^j}{\partial x^i}+
\frac{\partial A^i}{\partial x^j}\right)+\frac{1}c\frac{\partial\Phi}{\partial t}\delta_{ij}\;,\\
\label{pp5}
\Gamma^i_{00}&=&-
\frac{\partial\Phi}{\partial x^i}-\frac{4}c\frac{\partial A^i}{\partial t}\;,\\
\label{pp6}
\Gamma^i_{0j}&=&-2\left(
\frac{\partial A^i}{\partial x^j}-\frac{\partial A^j}{\partial x^i}\right)+\frac{1}c\frac{\partial\Phi}{\partial t}\delta_{ij}\;,\\
\label{pp7}
\Gamma^i_{jp}&=&-\delta_{jp}\frac{\partial\Phi}{\partial x^i}+\delta_{ip}\frac{\partial\Phi}{\partial x^j}+
\delta_{ij}\frac{\partial\Phi}{\partial x^p}\;,
\end{eqnarray}
One notices that our $\epsilon$-parametrization of the Einstein equations \cite{cqg} makes {\it all} time derivatives of the gravitational field vanish if the speed of gravity $c_g\rightarrow\infty$ ($\epsilon\rightarrow 0$). This makes sense since the limit $c_g\rightarrow\infty$ corresponds to the case of the instantaneous speed-of-gravity theory where the gravitational interaction is immediate from one place to another and both the gravity field equations and equations of motion of test particles, which depend on the affine connection (\ref{bb2})--(\ref{bb7}), do not contain any time derivative of the gravitational field. In contrast to this line of reasoning the PPN formalism insists that the equations of motion of test particles depend on the affine connection  (\ref{pp2})--(\ref{pp7}) that contains non-vanishing first time derivatives of the gravitational field in the limit $c_g\rightarrow\infty$. We consider this postulate as going against physical intuition. Furthermore, we emphasize that the magnitude of the time derivatives of the gravitational field in the affine connection has not been known as previous experiments in the solar system had been done in the static field approximation. The reader should also realize that in the PPN formalism the Lorentz transformation of the affine connection (\ref{pp2})--(\ref{pp7}) is identical with that for electromagentic field which is given by the matrix $\lambda^\alpha_\beta\equiv\Lambda^\alpha_\beta(\epsilon=1)$ depending on the speed of light $c$ alone. But the affine connection is a pure gravitational object having nothing in common with electromagnetic field and, hence, its Lorentz transformation are to be given by the matrix $\Lambda^\alpha_\beta(\epsilon)$, as it is adopted in our $\epsilon$-parametrized PN approach \footnote{The nature of gravity is more rich and is not reduced to that adopted in the metric-based theories of gravity -- the only class of the gravity theories discussed in \cite{will}. Actually, the affine connection is an independent geometric structure in more general class of the metric-affine gravity (MAG) theories \cite{hehl} making absolutely clear that the affine connection has no any association with electromagnetism.}.

 Double intergation of Eq. (\ref{lg}) along the unperturbed light ray yields the time of propagation of light from the point $x_0^i$ to $x^i_1=x^i(t_1)$
\begin{equation}
\label{qer}
t_1-t_0=\frac{1}{ c}|{\bm x}_1-{\bm x}_0|+\Delta(t_1,t_0)\;,
\end{equation}
where the relativistic time delay \cite{cqg}
\begin{equation}
\label{aa}
\Delta(t_1,t_0)=\frac{c}{ 2}\;k^\mu k^\nu k_\alpha\int_{t_0}^{t_1}dt\int_{-\infty}^t d\tau \Bigl[\Gamma^\alpha_{\mu\nu}(\tau,{\bm x})\Bigr]_{{\bm x}={\bm x}_N(\tau)}\;.
\end{equation}
Integration in Eq. (\ref{aa}) is along a straight line of unperturbed propagation of photon
\begin{equation}
\label{und}
{\bm x}_N(t)={\bm x}_0+c{\bm k}(t-t_0)\;,
\end{equation}
where $t_0$ is time of emission, $x^i_0$ is coordinate of the source of light at time $t_0$. Substituting equations (\ref{bb2})--(\ref{bb7}) to the time delay equation (\ref{aa}) and making use of the gauge condition (\ref{3}) in order to replace the time derivative of the potential $\Phi$ to the divergence of the potential ${\bm A}$, one recasts it into the following form
\begin{equation}
\label{ty}
\Delta(t_1,t_0)=2\int_{t_0}^{t_1}\Phi dt
-2\left(1-\frac{1}{\epsilon}\right)\int_{t_0}^{t_1}dt\int_{-\infty}^{t} {\bm\nabla}\cdot{\bm A}d\tau\;\;,
\end{equation}
where all terms unavailable to measurement with current VLBI technology are omitted, the potentials $\Phi$ and ${\bm A}$ are given by Eq. (\ref{8}) and are taken on the unperturbed light-ray trajectory (\ref{und}), that is ${\bm x}={\bm x}_N(t)$. 

Light is deflected the most strongly during a rather short interval of time when it passes close to the light-ray deflecting body \footnote{Precise determination of the size of the space region where the moving gravitational lens deflects light can be done if one estimates the magnitude of the relativistic deflection of light given by the acceleration-dependent terms in equation (\ref{pok}). Some research in this direction has been done in \cite{ks,k2,klk}.}. For this reason,  
integration of the relativistic time delay equation (\ref{ty}) can be performed with a good approximation under assumption of a uniform motion of the light-ray deflecting body (Jupiter, Sun) with constant velocity ${\bm v}$, that is 
\begin{equation}\label{pok}
{\bm z}(t)={\bm z}(t_{\rm A})+{\bm v}(t-t_{\rm A})+O(t-t_{\rm A})^2\;,
\end{equation}
where ${\bm z}(t_{\rm A})$ is position of Jupiter at time $t_{\rm A}$ and the acceleration-dependent terms have been neglected.  
The calculation is tedious but straighforward, and can be found in sections 3 and 4 of our paper \cite{cqg}. The result is given by \cite{cqg,pla}
\begin{eqnarray}
\label{tdel}
\Delta(t_1,t_0)&=&-\frac{2GM}{c^3}\ln\left(\frac{R_1-{\bm k}\cdot{\bm R}_1}{R_0-{\bm k}\cdot{\bm R}_0}\right)\;,
\end{eqnarray}
where ${\bm R}_1={\bm x}(t_1)-{\bm z}(s_1)$, ${\bm R}_0={\bm x}(t_0)-{\bm z}(s_0)$, $R_1=|{\bm R}_1|$,  $R_0=|{\bm R}_0|$, and the coordinates of the body in Eq. (\ref{tdel}) must be taken at the retarded time
\begin{eqnarray}
\label{a5}
s_1&=&t_1-\frac{\epsilon}{c}|{\bm x}_1-{\bm z}(s_1)|\;,\\\label{a6}
s_0&=&t_0-\frac{\epsilon}{c}|{\bm x}_0-{\bm z}(s_0)|\;.
\end{eqnarray}
Equations (\ref{a5}), (\ref{a6}) are null characteristics of the gravitational field equations (\ref{a1}), (\ref{a2}). Their appearance in the time delay equation (\ref{tdel}) is a direct consequence of the fact that the affine connection (\ref{bb2})--(\ref{bb7}) contains time derivatives which are normalized to the speed of gravity $c_g=c/\epsilon$ in accordance with the Lorentz-invariance properties of the gravitational field. In other words, the Lorenzt invariance of gravity and its finite speed of propagation are physically inseparable. This is the reason why the null characteristics of the gravitational field show up already in the linear terms beyond the static part of the Shapiro time delay. This part of the relativistic theory of gravitational field has been greatly misunderstood in \cite{asada,will-apj,carlip,s1,s2,pask,faber,will-page} who have missed the idea that the affine connection is a pure geometric object whose Lorentz transformation is determined by the value of the fundamental speed of gravity $c_g$ but not the speed of light $c$ that defines the Lorentz transformation of electromagentic field alone \footnote{It is also possible to look at this problem from the point of view of the affine-metric theories of gravity \cite{hehl}. If the fundamental speed of gravity $c_g\not= c$ then the affine connection suggests the presence of {\it nonmetricity} different from zero. A nonvanishing nonmetricity assumes violation of the Lorentz invariance for gravity. Researchers in \cite{asada,will-apj,carlip,s1,s2,pask,faber,will-page} postulate that the nonmetricity is zero but this assumption is a matter of experimental testing but not a theoretical belief.}.   

Equations (\ref{tdel}), (\ref{a5}) were used in the Jovian light deflection experiment in order to measure the speed-of-gravity parameter $\epsilon$ \cite{fk}. The goal of the October 2005 VLBI experiment is to confirm that the retarded time equation (\ref{a5}) is valid with accuracy about 1 \%. Before discussing the October 2005 experiment we focus on the interpretation of the Jovian deflection experiment \cite{k1,fk} to clear out gravitational physics of the relativistic time-delay measurements in the field of moving gravitational lenses (see also \cite{cqg,frit}).  

\section{Physical Interpretation of the Jovian Deflection Experiment}

\subsection{Relativistic Delay Measurement as a Null-Type Experiment}

Equation (\ref{tdel}) describes relativistic time delay of light (radio wave) caused by a massive body moving with constant velocity ${\bm v}$ with respect to the rest frame of observer located at the point ${\bm x}$. Gravity is a long-ranged field and the time delay given by Eq. (\ref{ty}) is effectively an integral effect of the gravitational force exerted on the photon along its entire trajectory. However, the strongest impact of the time-dependent gravitational field on the photon emitted at time $t_0$ is when the massive body is located in its retarded position ${\bm z}(s_1)$ taken at the retarded time $s_1$ determined by Eq. (\ref{a5}). If general relativity were correct, then, one had $c_g=c$ and only the gravitoelectric potential $\Phi$ of the body would be essential for calculation of the time delay (\ref{tdel}). However, if $c_g\not=c$ $(\epsilon\not=1)$ the gravitomagnetic potential ${\bm A}$ gives additional contribution to the delay as shown by Eq. (\ref{ty}). This term is to vanish in general relativity. Therefore, measuring divergence of parameter $\epsilon$ from its general relativistic value, $\epsilon=1$, is a null test of general relativity. This idea was not captured in \cite{s1,s2} where equality $\epsilon=1$ is accepted as a postulate not requiring experimental verification.

\subsection{Measurement of the Gravitomagnetic Field}

The second term (the double integral) in the right side of Eq. (\ref{ty}) is caused by the gravitomagnetic field due to the orbital motion of the light-ray deflecting body (Jupiter, Sun). This term leads to dependence of the retarded time (\ref{a5}) on the parameter $\epsilon$. If general relativity is correct the contribution of the double integral in Eq. (\ref{ty}) to the time delay is identically zero, and the magnitude of the gravitomagnetic field is given by Eqs. (\ref{8}), (\ref{8a}) with $\epsilon=1$. By observing the relativistic deflection of light by moving Jupiter we confirmed \cite{fk} that $\epsilon=1$ with precision 20\% which means that the {\it extrinsic} gravitomagnetic field given by Eq. (\ref{8}) really exists and works as predicted by general relativity. This point has been misinterpreted by Pascual-Sanchez \cite{pask} who confused the effect of the relativistic gravitomagnetic field with measurment of the classic R\"omer delay.

\subsection{The Lorentz Invariance of Gravity}

Parameter $\epsilon$ signifies the Lorentz transformation of gravitational field variables, makes its effect clearly visible in theoretical calculations, and distinguishes the fundamental speed of gravity $c_g$ from the speed of light $c$ because the matrix of the Lorentz transformation of the gravitational field variables, $\Lambda^\alpha_{\;\beta}(\epsilon)$, depends on the speed of gravity $c_g=c/\epsilon$. If $\epsilon\not=1$, the Lorentz invariance of gravity is broken with respect to the Lorentz invariance of electromagentic field. Thus, measuring $\epsilon$ allows us to test the compatibility of the Maxwell and Einstein equations. Jovian experiment confirmed that they are compatible as predicted by general relativity with 20\% of accuracy.

\subsection{The Aberration and the Speed of Gravity}

The $\epsilon$-parameterization of the Einstein equations with the single parameter $\epsilon$ helps to keep track of any relativistic effect associated with the fundamental upper limit on the speed of gravity $c_g=c/\epsilon$.  
Physically, if the fundamental upper limit on the speed of gravity $c_g=\infty$ all time derivatives of the gravitational field would vanish and, hence, could not be detected. Any gravitomagnetic effect would be completely supressed as well. However, the Jovian experiment confirmed that $\epsilon=1$ thus revealing that the time derivatives of the metric tensor in the affine connection exist and give contribution to the relativistic deflection of light in the full agreement with general relativity. Non-broken Lorentz invariance of gravity confirmed in the Jovian experiment \cite{fk} also means that its speed of propagation is finite \footnote{Jovian experiment sets a rather stringent upper limit on the speed $c_{\rm gw}$ of gravitational waves since in general relativity and other valid gravitational theories $c_{\rm gw}\le c_g$ as proven in \cite{bs}.}. This interpretation is in a full agreement with the causal (retarded) nature of gravity which is revealed in Eqs. (\ref{a5}), (\ref{a6}) describing the null characteristics of the gravity field equations (\ref{a1}), (\ref{a2}) connecting the field point ${\bm x}_1$ (${\bm x}_0$)  with the retarded position ${\bm z}(s_1)$ (${\bm z}(s_0)$) of the light-ray deflecting body (Jupiter). The null characteristics of gravity can not be confused with the null characteristics of light. Indeed, light propagates from a source of light (quasar) to observer while, for example, the null vector ${\bm N}_1={\bm R}_1/R_1$, where ${\bm R}_1={\bm x}_1-{\bm z}(s_1)$, points from observer, located at ${\bm x}_1$, to the gravitating body (Jupiter) located at ${\bm z}(s_1)$ along the null cone of propagation of gravity. We measured \cite {fk} the retarded pozition of Jupiter ${\bm z}(s_1)$ corresponding to the time of observation $t_1$ by two independent methods: (1) using radio waves transmitted from the spacecrafts orbiting Jupiter, (2) using the gravitational time delay of light of the quasar. The radio-tracking retarded position of Jupiter depends on the speed of light $c$ and can be read out of the JPL ephemeris \cite{standish}. On the other hand, the retarded position ${\bm z}(s_1)$ of Jupiter in the gravitational time delay (\ref{tdel}) depends on the speed of gravity $c_g=c/\epsilon$ and can be determined independently of JPL ephemeris from precise measurement of the relativistic time delay by VLBI.  
We have proved \cite{fk} that the time-delay retarded position of Jupiter due to the speed of gravity $c_g$ is in accordance with its retarded position obtained independently from the radio tracking of spacecrafts for $\epsilon=1$ which means that gravity does not propagate faster than light. Briefly, the Jovian experiment measured the retardation of gravity effect given by equation (\ref{a5}) with respect to the light travel time which was read off from the JPL ephemeris of Jupiter.

The null-type character of the Jovian experiment closely associated with the test of the Lorentz invariance of the gravitational field has been overlooked by a number of researchers who claimed that the Jovian deflection expeirment measured the aberration (speed) of light \cite{asada,will-apj,carlip,s1,s2,pask}. This claim is erroneous since the aberration of light is special relativistic effect in flat space-time when gravity is absent. The Jovian experiment, however, studies propagation of light in curved space-time caused by the gravitational field of Jupiter and Sun. It measured the retarded position of Jupiter on its orbit not by observing Jupiter's own radio emission, as mistakenly assumed in \cite{asada,s1,s2}, but from ultra-precise measurement of the direction of the gravitational force exerted by moving Jupiter on the quasar's photons in the plane of the sky \cite{k1,fk,cqg}. This result can be obtained theoretically in two independent ways: (1) by making use of the retarded Lienard-Wiechert potentials of Eqs. (\ref{a1})-(\ref{a2}) \cite{ks,k1}, (2) by making use of the Lorentz transformation technique \cite{klion}. The two methods lead to the same results given by Eqs. (\ref{tdel}), (\ref{a5}), (\ref{a6}). The Lorentz transformation of the time delay from static to a moving frame means that both light and gravity field variables entering Eq.  (\ref{aa}) must be transformed simultaneously. The aberration of light transforms vector $k^\alpha=(1,{\bm k})$ of the light ray propagation alone. If only the aberration of light is taken into account in equation (\ref{aa}) without transforming the affine connection (gravitational force) the time delay can not maintain its invariance and the terms proportional to $\delta=c/c_g-1$ will emerge. If the Lorentz transformation matrix of the affine connection is the same as for light, then, $\delta=0$ and the time delay is Lorentz-invariant with the position of the light-ray deflecting body taken at the retarded time $s_1$ as given by Eq. (\ref{a5}) with $\epsilon=1$. The Jovian experiment used the aberration of light exclusively as a calibrating standard whose characteristics (defined by the speed of light $c$) are known from laboratory measurements. The speed of gravity $c_g$ was measured with respect to this calibrating standard (the aberration of light effect) confirming that the Lorentz transformation matrix of the gravitational field (the aberration of gravity effect) is the same as that for light \cite{comment}. 

\section{Measuring the Aberration of Gravity with the Quasar-Sun Encounter} 

Accuracy in measuring the aberration of gravity effect can be significantly improved by observing gravitational deflection of light on the limb of Jupiter and/or Saturn by space missions SIM and/or GAIA. Another method is based on measuring quasar's gravitational time delay by the Sun directly in the geocentric reference frame. All previous measurements of this effect had been done in the barycentric frame of the solar system with respect to which the Sun is almost static and the dynamical effects caused by time variations of its gravitational field are negligibly small. However, Sun is a massive gravitational lens moving with respect to the geocentric frame and it must deflect light in this frame from its retarded position defined by Eq. (\ref{a5}) with $\epsilon=1$ if general relativity is valid. If the aberration of gravity is not taken into account (that is, $\epsilon=0$) it will cause a small displacement $\delta\alpha$ of the quasar's position in the geocentric frame from that given by Einstein's formula $\alpha=4GM_\odot/(c^2d)$, where $M_\odot$ is mass of the Sun, and $d$ is the impact parameter of the light ray with respect to the Sun taken at the retarded instant of time. The aberration of gravity effect is estimated by Eq. \cite{k1} 
\begin{equation}
\label{ura}
\delta\alpha=\alpha_\odot\left(\frac{1 {\rm AU}}{R_\odot}\right)\left(\frac{R_\odot}{d}
\right)^2\left(\frac{v_\odot}{c}\right)\;,
\end{equation}
where $\alpha_\odot=1.75"$ is the light deflection on the solar limb, 1 {\rm AU} is one astronomical unit, $R_\odot=7\times 10^{10}$ cm is radius of the Sun, and $v_\odot$ is velocity of the Sun relative to the Earth, which is about the barycentric velocity of the Earth, $v_E\simeq 30$ km/s. Substituting the numbers one obtains
\begin{equation}
\label{ura2}
\delta\alpha=37.5\left(\frac{R_\odot}{d}\right)^2 \mbox{mas}\;. 
\end{equation}
In October 2005 quasar 3C279 will pass by the Sun at the distance of 2 solar radii \cite{fomalont}. If one takes the minimal value of the impact parameter allowed by the solar corona  as $d\simeq 4 R_\odot$, it yields for the aberration of gravity effect the maximal magnitude of $\delta\alpha\simeq 2344$ $\mu$arcseconds. Assuming that the precision of VLBI measurement of the quasar's position is 20 $\mu$arcseconds we shall be able to measure the aberration of gravity effect (parameter $\epsilon=c/c_g$) with the accuracy approaching to 1\% as contrasted to 20\% in the case of the Jovian deflection experiment \cite{fk}. 

We notice that the data from the recent Cassini mission experiment by Bertotti, Iess and Tortora \cite{bi} can be used for measuring the aberration of gravity effect and setting the upper limit on the violation of the Lorenzt invariance of gravity with accuracy presumably approaching to 0.5 \%. Space-based experiments like LATOR \cite{lat} and/or ASTROD \cite{astrod} will help us to reduce this limit down to $10^{-4}$ \footnote{In this connection we would like to emphasize that the magnitude of the aberration of gravity effect given in \cite{will-page} is over-pessimistic. Realistic estimate is given in \cite{fomalont}.}.

\section*{Acknowledgments}
We wish to acknowledge the grant support of the University of Missouri-Columbia and the Eppley Foundation for Research. We thank G.E. Melki for valuable comments
and an anonymous referee for instructive
report.


\begin{thebibliography}{99}
\bibitem{bct}V.~B. Braginskii,  C.~M.
Caves \& K.~S. Thorne,\ (1977), {\it Phys. Rev. D}, {\bf 15}, 2047 
\bibitem{m1} B. Mashhoon,\ (2001), In: 
{\it Reference Frames and Gravitomagnetism}, Proceedings of the XXIII Spanish Relativity Meeting held 6-9 September, 2000 in Valladolid, Spain. Eds. J.F. Pascual-Sánchez, L. Floriá, A. San Miguel and F. Vicente. World Scientific Publising Company, 2001., p.121 
\bibitem{rugt}M.~L. Ruggiero \& 
\& A. Tartaglia, \ (2002), {\it N. Cim. B}, {\bf 117}, 743 
\bibitem{mht} B. Mashhoon, F.~W. Hehl 
 \& D.~S. Theiss, \ (1984), {\it Gen. Rel. Grav.}, {\bf 16}, 711
\bibitem{ashall} N. Ashby \& T. Allison, \ (1993), {\it Cel. Mech. Dyn. Astron.}, {\bf 57}, 537
\bibitem{ior} L. Iorio, \ (2001), {\it N. Cim. B}, {\bf 116}, 777    
\bibitem{cw} I. Ciufolini \&  J.A. Wheeler, \ (1995), {\it Gravitation and Inertia} (Princeton: Princeton University Press, 1995)
\bibitem{ashss} N. Ashby \& B. Shahid-Saless, \ (1990), {\it Phys. Rev. D}, {\bf 42}, 1118
\bibitem{gpb} C.W.F. Everitt, S. Buchman, D.B. DeBra, G.M. Keiser, J.M. Lockhart, B. Muhlfelder, B.W. Parkinson, J.P. Turneaure \& and other members of the Gravity Probe B team, \ (2001) In: {\it Gyros, Clocks, Interferometers...: Testing Relativistic Gravity in Space}, Eds. C. L\"ammerzahl, F. Everitt and F.W. Hehl. Springer-Verlag: Berlin, 2001) pp. 52--82
\bibitem{gpb1} See http://einstein.stanford.edu/
\bibitem{iormor} L. Iorio \& A. Morea, \ (2004), {\it Gen. Rel. Grav.}, {\bf 36}, 1321
\bibitem{cp}I. Ciufolini \& E.~C. Pavlis, \ (2004), {\it Nature}, {\bf 31}, 958 
\bibitem{ior2} L. Iorio, \ (2005), {\it New Astron.}, {\bf 10}, 603
\bibitem{ior3} L. Iorio, \ (2005), gr-qc/0507041
\bibitem{ior4} L. Iorio \& V. Lainey, \ (2005), {\it Int. J. Mod. Phys. D}, in press (eprint gr-qc/0508112)
\bibitem{ein} A. Einstein, \ (1911), Ann. Phys., {\bf 35}, 898
\bibitem{shapiro} I.I. Shapiro, \ (1964),  {\it Phys. Rev. Lett.}, {\bf 13}, 789
\bibitem{ks}  S.M. Kopeikin \& G. Sch\"afer, \ (1999), {\it Phys. Rev D}, {\bf 60}, 124002 
\bibitem{km} S.M. Kopeikin \& B.Mashhoon, \ (2002), {\it Phys. Rev D}, {\bf 65}, 064025
\bibitem{LL}  L.D. Landau \& E.M. Lifshitz, \ (1971), {\it The Classical Theory of Fields} (Oxford: Pergamon) 
\bibitem{mtw} C.W. Misner, K.S. Thorne and J.A. Wheeler, (1973), {\it Gravitation} (New York: Freeman, 1973)
\bibitem{fs} T. Futamase \& B.F. Schutz, \ (1983), {\it Phys. Rev D}, {\bf 28}, 2363 
\bibitem{cqg} S.M. Kopeikin,  \ (2004), {\it Class. Quantum Grav.}, {\bf 21}, 3251
\bibitem{will} C.~M. Will, \ (1993), {\it Theory and experiment in gravitational physics} (Cambridge: Cambridge University Press)
\bibitem{will-r} C.~M. Will, \ (2001) {\it The Confrontation between General Relativity and Experiment}, Living Rev. Relativity {\bf 4}, 4. URL (cited on July 16, 2005):
http://www.livingreviews.org/lrr-2001-4 
\bibitem{k1} S.M. Kopeikin, \ (2001), {\it Astrophys. J.} Lett., {\bf 556}, L1
\bibitem{fk}  E.B. Fomalont \& S.M. Kopeikin, \ (2003), {\it Astrophys. J.}, {\bf 598}, 704
\bibitem{ckmr} I. Ciufolini, S. Kopeikin, B. Mashhoon \& F. Ricci, \ (2003), {\it Phys Lett. A}, {\bf 308}, 101
\bibitem{tara1} A. Tartaglia 
\&  M.~L. Ruggiero, \ (2004), {\it Gen. Rel. Grav.}, {\bf 36}, 293 
\bibitem{ser} M. Sereno, \ (2005), {\it Mon. Not. R. Astron. Soc.}, {\bf 357}, 1205
\bibitem{lat} S.~G. Turyshev, M. Shao \& K. Nordtvedt, \ (2004), {\it Astronom. Nachrichten}, {\bf 325}, 267 
\bibitem{astrod}W.-T. Ni, S. Shiomi \&  A.-C. Liao, \ (2004), {\it Class. Quant. Grav.}, {\bf 21}, 641 
\bibitem{tr1} A. Tartaglia,  M.~L.
Ruggiero \& A. Nagar, \ (2005), {\it Phys. Rev. D}, {\bf 71}, id. 023003 
\bibitem{tr2} L. M. 
Ruggiero \& A. Tartaglia, \ (2005), e-print: gr-qc/0509098 
\bibitem{dk} O.V. Doroshenko \& S.M. Kopeikin, \ (1995), {\it Mon. Not. R. Astron. Soc.}, {\bf 274}, 1029 
\bibitem{wk} N. Wex \& S.M. Kopeikin, \ (1999), {\it Astrophys. J.}, {\bf 514}, 388 
 \bibitem{jackson} J.D. Jackson, (1998), {\it Classical Electrodynamics}, (New York: John Wiley \& Sons, Inc., 1998)
 \bibitem{hehl} C. Heinicke, P. Baekler \& F.W. Hehl, \ (2005), {\it Phys. Rev. D}, {\bf 72}, id. 025012

\bibitem{k2}S.M. Kopeikin, \ (1990), {\it Soviet Astronomy}, {\bf 34}, 5 
\bibitem{klk} S.A. Klioner \& S.M. Kopeikin, \ 1992, {\it Astron. J.}, {\bf 104}, 897 
\bibitem{pla} S.M. Kopeikin, \ (2003), Phys. Lett. A., {\bf 312}, 147
\bibitem{asada} H. Asada, \ (2002), {\it Astrophys. J. Lett.}, {\bf 574}, L69 
\bibitem{will-apj} C.M. Will, \ (2003), {\it Astrophys. J.}, {\bf 590}, 683
\bibitem{carlip}  S. Carlip S., \ (2004), {\it Class. Quant. Grav.}, {\bf 21}, 3803
\bibitem{s1} S. Samuel, \ (2003), {\it Phys. Rev. Lett.}, {\bf 90}, 231101
\bibitem{s2} S. Samuel, \ (2004), Int. J. Mod. Phys. D, {\bf 13}, 1753
\bibitem{pask} J.-F. Pascual-Sanchez, \ (2004), Int. J. Mod. Phys., {\bf 13}, 2345
\bibitem{faber} J. Faber, \ (2003), e-print: astro-ph/0303346 (unpublished)
\bibitem{will-page} See http://wugrav.wustl.edu/people/CMW/SpeedofGravity.html
\bibitem{frit} S. Frittelli, \ (2003), {\it Mon. Not. R. Astron. Soc.}, {\bf 344}, L85 
 \bibitem{bs} M.\'A.G. Bonilla \& J.M.M. Senovilla, \ (1997), {\it Phys. Rev. Lett.}, {\bf 78}, 783
\bibitem{standish} Standish E. M. 2003 {\it JPL planetary ephemeris DE410} Interoffice Memorandum 312.N-03-109 
\bibitem{klion} S.A. Klioner, \ (2003), Astron. Astrophys., {\bf 404}, 783
\bibitem{comment} S.M. Kopeikin, \ (2005), {\it Class. Quant. Grav.}, in press; e-print:gr-qc/0510048
\bibitem{fomalont} S.M. Kopeikin \& E.B. Fomalont, \ (2005), e-print:gr-qc/0510077 
\bibitem{bi} B. Bertotti, L. Iess \& P. Tortora, \ (2003), {\it Nature}, {\bf 425}, 374 
 
 
\end{thebibliography}
\end{document}